\documentclass[12pt,a4paper]{article}
\usepackage{graphics}
\begin{document}
\baselineskip 0.25in
\title{\Huge{How to Run Mathematica Batch-files in Background ? }}
\author{\Large {Santanu K. Maiti} \\ \\
         \Large {E-mail: {\em santanu.maiti@saha.ac.in}} \\ \\
          \Large {Theoretical Condensed Matter Physics Division} \\
           \Large {Saha Institute of Nuclear Physics} \\
            \Large {1/AF, Bidhannagar, Kolkata-700 064, India}}
\date{}
\maketitle
\newpage
\tableofcontents

\newpage
\begin{center}
\addcontentsline{toc}{section}{\bf {Preface}}
{\Large \bf Preface}
\end{center}
Mathematica is a versatile equipment for doing numeric and symbolic
computations and it has wide spread applications in all branches of 
science. Mathematica has a complete consistency to design it at every 
stage that gives it multilevel capability and helps advanced usage 
evolve naturally. Mathematica functions work for any precision of number 
and it can be easily computed with symbols, represented graphically to 
get the best answer. Mathematica is a robust software development that 
can be used in any popular operating systems and it can be communicated 
with external programs by using proper mathlink commands. 

Sometimes it is quite desirable to run jobs in background of a computer
which can take considerable amount of time to finish, and this allows 
us to do work on other tasks, while keeping the jobs running. Most of us
are very familiar to run jobs in background for the programs written
in the languages like C, C$++$, F77, F90, F95, etc. But the way
of running jobs, written in a mathematica notebook, in background is 
quite different from the conventional method. In this article, we
explore how to create a mathematica batch-file from a mathematica 
notebook and run it in background. Here we concentrate our study
only for the Unix version, but one can run mathematica programs in 
background for the Windows version as well by using proper mathematica
batch-file.

\newpage
\section{Introduction}

Mathematica is the world's best powerful computing system which has 
been released in 1988 and have profound evidences on the way of
computations that are used in technical and other fields of work. The key
intellectual aspect in mathematica is the invention of a new kind of
symbolic computation language that can manipulate the very wide range of 
objects needed to achieve the generality required for technical computing
by using a very small number of basic primitives. Mathematica has wide 
spread applications in every branch of sciences-physical, biological,
social, and other and has played a crucial role in many important 
discoveries and has been basis for thousands of technical papers.

Mathematica is a versatile and powerful package for calculating 
mathematics and publishing mathematical results. It can be used in almost 
all popular workstation operating systems like, Microsoft Windows, Apple
Macintosh operating system, Linux and other Unix-based systems. As a
programming tool Mathematica~\cite{wolfram} provides a rich set of 
programming extensions. Programming can be done in different ways like 
functional, logical i.e., rule-based or procedural type or a mixture of 
all these three. The another most important aspect is that, mathematica 
provides mathlink~\cite{maeder,san} which allows mathematica programs 
to communicate with external programs written in C, Java, XL-Fortran 
languages or any other languages. Mathematica is now emerging as an 
important tool in many branches of computing, and today it stands as 
the world's best system for general computation.

In the previous article~\cite{san}, we have studied in detail how to start 
mathematica, write programs in mathematica and the way of linking of 
external programs with a mathematica notebook by using proper mathlink
commands. Now it is quite desirable to run jobs in background which take
much time to finish and to do other works in separate windows, keeping the 
jobs running. This motivates us to explore the basic mechanisms for running
mathematica programs in background. It can be done by creating proper 
mathematica batch-file which we will describe here elaborately, and this 
study may be quite helpful for us.

\section{How to Create a Mathematica Batch-file and Run it in Background}

In order to understand the complete process, let us start by giving a very 
simple example of a mathematica program. We set the program as follows: \\
{\em The generation of a list of two random numbers, a $2$D plot from 
these set of random numbers and then the creation of an `EPS' file for 
this 2D plot}.

\vskip 0.12in
\noindent
For this program, first we need to make a list of two random numbers
and then construct a 2D plot using this set of random numbers. Finally,
we make an `EPS' file for this plot. Our main aim of this article is
to run this complete job in background. Before doing this job
in background, let us now describe the different mathematical operations 
with proper commands which are to be done in a mathematica notebook for 
this particular job.

\vskip 0.1in
\noindent
The program for the generation of a set of two random numbers and a
2D plot from these numbers is as follows:

\vskip 0.2in
\begin{center}
{\fbox{\parbox{5.65in}{\centering{
sample$[$times$_-]$$:=$Block$[\{$local variables$\}$,

numbers$=$Table$[\{$Random$[]$, Random$[]$$\}$, $\{i$, $1$, times$\}]$;

fig$=$ListPlot$[$numbers, PlotJoined$\rightarrow$True,
AxesLabel$\rightarrow$$\{$xlabel, ylabel$\}$$]]$
}}}}
\end{center}

\vskip 0.12in
\noindent
To get the output of this program, we run it by entering some value for 
the variable `times', like `sample[100]' or `sample[200]' etc. Then the
\begin{figure}[h]
{\centering \resizebox*{10.0cm}{6.5cm}{\includegraphics{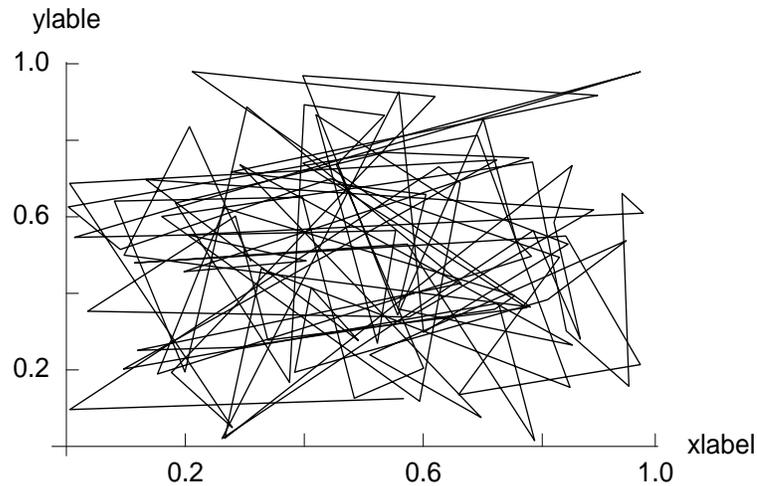}}\par}
\caption{A $2$D plot for a set of two random numbers.}
\label{random}
\end{figure}
mathematica does the proper operations and executes the result in an 
output cell. The output of the 2D plot is shown in Fig.~\ref{random}. 

\vskip 0.12in
\noindent
Now to create an `EPS' file for this 2D plot we use the following 
operation:

\vskip 0.2in
\begin{center}
{\fbox{\parbox{3.15in}{\centering{
Export$[$``filename.eps", fig, ``EPS"$]$
}}}}
\end{center}  

\vskip 0.12in
\noindent
In this above expression, the name `fig' is used to call the graphics 
file, and the `eps' file is saved by the name `filename.eps' in the 
present working directory.

Thus we are now clear about all the mathematical operations those
are to be done in a mathematica notebook for the above mentioned program. 
So now we make our attention for running this program in background. 

In order to run this program in background, first we need to create a
batch-file which is a text file from these mathematica input commands 
those are written in different cells of a mathematica notebook. For this
purpose, we go through these steps:

(a) Select the cells from the mathematica notebook, and then follow the 
direction by clicking on 
{\em Cell} $\rightarrow$ 
{\em Cell Properties} $\rightarrow$ {\em Initialization Cell} from the menu 
bar to initialize the cells. 

(b) To generate the batch-file, follow the direction by clicking on 
{\em File} $\rightarrow$ {\em Save As Special\ldots} $\rightarrow$
{\em Package Format} from the menu bar. 

Then a dialog box appears for specifying the file name and the location
of the mathematica input file. Here we use the input file for the 
operation of the mathematica job.

After these steps, let us suppose, we generate a batch-file named as
`santanu.m' for the above mathematica program. Generally the batch-files 
for this purpose are specified by using the extension `.m' i.e., like 
the name as `filename.m'. To run this batch-file `santanu.m' in background, 
we use the following prescription:

\vskip 0.2in
\begin{center}
{\fbox{\parbox{4in}{\centering{
nohup time math $<$ santanu.m $>$ santanu.out $\&$
}}}}
\end{center}

The file name `santanu.out' is the output file, where all the outputs
for the different operations are available. To get both the input and 
output lines of the mathematica notebook, it is necessary to use the 
following command in the first line of the notebook.

\vskip 0.2in
\begin{center}
{\fbox{\parbox{2.35in}{\centering{
AppendTo[$\$$Echo, ``stdout"]
}}}}
\end{center}

At the end of all these steps, we get the output file `santanu.out' and
the graphics file `filename.eps' in `EPS' format in the present working
directory where the batch-file `santanu.m' is run in the background. 

\newpage
\noindent
\addcontentsline{toc}{section}{\bf {Concluding Remarks}}
\begin{flushleft}
{\Large \bf {Concluding Remarks}}
\end{flushleft}
\vskip 0.1in
\noindent
In summary, we have addressed in detail how to set up a mathematica 
batch-file from a mathematica notebook and run it in the background 
of a computer. Several programs are there which can take a considerable 
amount of time to run. Some may take few days or even few weeks to 
complete their analysis. For this reason, it may be desirable to place 
such jobs in the background. This is a way of running a program that 
allows one to continue working on other tasks (or even log out) while 
still keeping the program running. Furthermore, backgrounded jobs are 
not dependent on our session remaining open, so even if our computer 
crashes, the job will continue uninterrupted.

\noindent
\addcontentsline{toc}{section}{\bf {Acknowledgment}}
\begin{flushleft}
{\Large \bf {Acknowledgment}}
\end{flushleft}
\vskip 0.1in
\noindent
I acknowledge with deep sense of gratitude the illuminating comments and
suggestions I have received from Prof. Sachindra Nath Karmakar during the
preparation of this article.

\addcontentsline{toc}{section}{\bf {References}}

\end{document}